\documentclass[aps,prl,twocolumn,showpacs,preprintnumbers,amsmath,amssymb,
epsf,superscriptaddress,groupedaddress, tightenlines]{revtex4}

\usepackage{epsfig,graphicx}
\usepackage{amsmath}
\usepackage{bm}
\newcommand{\nn}{\nonumber}
\newcommand{\vslash}{v\hspace*{-5.5pt}\slash}
\newcommand{\Aslash}{{\cal A}\hspace*{-5.5pt}\slash}
\newcommand{\Bslash}{{\cal B}\hspace*{-5.5pt}\slash}

\newcommand{\varepsslash}{\varepsilon\hspace*{-5.5pt}\slash}
\newcommand{\lqcd}{\Lambda}
\def\nslash{n\!\!\!\slash}
\def\bnslash{\bar n\!\!\!\slash}
\def\bn{\bar n}
\newcommand{\ra}[1]{{\bf  #1}}

\begin{document}
\preprint{UCSD/PTH 05-12}
\preprint{MIT-CTP 3674}


\title{
The CP asymmetry in $B^0(t) \to K_S\pi^0\gamma$ in the Standard Model}


\author{Benjam\'\i{}n Grinstein}
\affiliation{Department of Physics, University of California at San Diego,
  La Jolla, CA 92093}

\author{Dan Pirjol}
\affiliation{Center for Theoretical Physics, Massachusetts Institute for
  Technology, Cambridge, MA 02139}


\date{\today}

\begin{abstract}
The time-dependent CP {\ra asymmetry} in exclusive $B^0(t)\to K^{*0}\gamma$ decays 
has been proposed as a probe of new physics in B decays. Recently, this method 
was extended
to radiative decays into multibody hadronic final states such as 
$B^0(t)\to K_S\pi^0\gamma$ and $B^0(t)\to \pi^+\pi^-\gamma$.
The CP asymmetry in these decays vanishes to the extent that the
photon is completely polarized. In the Standard Model, the photon
emitted in $b\to s\gamma$ has high left-handed polarization, but right-handed
contamination enters already at leading order in $\lqcd/m_b$ even
for vanishing light quark masses. We compute here the magnitude of this
effect and the time dependent CP asymmetry parameter $S_{K_S \pi^0\gamma}$.
We find that the Standard Model can easily accomodate values of $S$ as large
as $10\%$, but a precise value cannot be obtained at present
because of strong interactions uncertainties.
\end{abstract}



\maketitle


\section{I. Introduction}

The standard model (SM) predicts that photons are predominantly left-handed in 
$b\to q\gamma$ ($q = s, d$) decay (and right-handed in $\bar b\to \bar q\gamma$).  
In the presence of new physics this prediction can be changed, and a significant
right-handed photon amplitude can appear in $b\to s\gamma$ decays.
Several methods have been suggested for testing this prediction in radiative
B decays~\cite{Atwood:1997zr,Grossman:2000rk}.

One of these methods makes use of time-dependent CP violation in 
$B^0\to f\gamma$ with $f$ a CP eigenstate~\cite{Atwood:1997zr}. 
Since $\gamma_L$ and
$\gamma_R$ cannot interfere, the time dependent $CP$ asymmetry
\begin{eqnarray}\label{ACPt}
&&{} {\Gamma[\bar B^0(t)\to f \gamma] - \Gamma[B^0(t)\to f\gamma]\over
   \Gamma[\bar B^0(t)\to f\gamma] + \Gamma[B^0(t)\to f\gamma] } \nn\\
&&{}\qquad = S_{f\gamma} \sin(\Delta m\, t)
   - C_{f\gamma} \cos(\Delta m\, t)\,,
\end{eqnarray}
is sensitive to the ratio of the right-/left-handed photon amplitudes.  
These can be parameterized as
\begin{eqnarray}
r_f e^{i\delta_f} = \eta_{\rm CP}(f) \frac{A(\bar B^0\to f\gamma_R)}
{A(\bar B^0\to f\gamma_L)}
\end{eqnarray}
where $\eta_{\rm CP}(f)$ is the CP eigenvalue of the state $f$. 
This method has been extended in Ref.~\cite{Atwood:2004jj} also to 
decays into multibody final states, such as for example $B\to K_S\pi^0\gamma$.

Summing over the unobserved photon polarization, the CP violating parameter 
in Eq.~(\ref{ACPt}) is given by
\begin{equation}\label{Ss}
S_{f\gamma} = - \frac{2r_f}{1+r_f^2} \cos\delta_f \sin2\beta \,.
\end{equation}
In the SM, it is usually assumed (incorrectly, see \cite{pol}) 
that $r_q \sim m_q/m_b$, which leads
to a small CP asymmetry $S \sim 2\%$
in the $b\to s\gamma$ transition. We used in this estimate $\sin 2\beta =
0.685 \pm 0.032$ \cite{hfag}.

Measurements of the $CP$ asymmetries in $B\to K^*\gamma$ were reported 
by  BABAR~\cite{Aubert:2005bu} and
BELLE~~\cite{Ushiroda:2005sb}, see Table I.
In addition, these two collaborations measured the $CP$ asymmetry 
in the nonresonant mode
$S_{K_S\pi^0\gamma}$ in two different ways: a) BABAR excludes the 
$K^*$ resonance  by integrating over the $K_S \pi^0$ invariant mass range
1.1 GeV $< M_{K_S\pi^0} < 1.8$\, GeV; b) BELLE includes both resonant
and nonresonant modes by integrating over the range
$m_K + m_\pi \leq M_{K_S\pi^0} \leq 1.8$ GeV.  The error in these
determinations is still too large to allow a meaningful comparison with the
SM prediction. 
With a view to improving the statistics of such measurements, we would like to 
assess the feasibility of combining the resonant and
nonresonant measurements.
Also, it is clear that searching for new physics with such 
measurements requires a reliable estimate of the standard model background.

\begin{table}[b]\label{table1}
  \begin{center}
    \begin{tabular}{|c|cc|}
\hline
 $f$  &  \multicolumn{2}{|c|}{$S_f$}  \\
\cline{2-3} 
 & BABAR & BELLE \\
\hline
\hline
$K^{*0}\gamma$ & $-0.21 \pm 0.40 \pm 0.05$ & $+0.01\pm 0.52\pm 0.11$ \\
$[K_S\pi^0]_{\rm nonres}\gamma$ & $0.9 \pm 1.0 \pm 0.2$ & 
   $+0.20 \pm 0.66$ \\
$K_S\pi^0\gamma$ & $-0.06\pm 0.37$ & $+0.08 \pm 0.41 \pm 0.10$ \\
\hline
\end{tabular}
\end{center}
{\caption{Experimental results for the CP asymmetry parameter $S_f$ for
$f=K^{*0}\gamma$ and $f=K_S \pi^0\gamma$ from BABAR and BELLE. The BABAR
nonresonant region includes all states with 
1.1\ GeV $< m_{K_S\pi^0} < 1.8$ GeV, while BELLE uses the range
1.0\ GeV $< m_{K_S\pi^0} < 1.8$ GeV. The errors shown are statistical and 
systematic, respectively.}}
\end{table}

At leading order in $1/m_b$, the photon
emitted in $B\to K^*\gamma$ is always left-handed polarized, to all orders
in $\alpha_s$ \cite{pol}. However, in multibody decays such as $B\to K_S \pi^0\gamma$
a right-handed component appears, already at leading order in $1/m_b$, in the kinematical
region with an energetic kaon and a soft pion. This is mediated by a
$B^*$ pole diagram, with the $B^*\to K\gamma_R$ amplitude 
calculable at leading order in $\lqcd/m_b$ using factorization in QCD
as following from Soft Collinear Effective Theory (SCET)  \cite{pol}. 
The presence of two hadrons in the final state evades the helicity argument 
which forbids a right-handed photon in $\bar B\to \bar K^*\gamma$. Note that 
a right-handed photon  is present
in inclusive $B \to X_s\gamma$ at leading order \cite{pol}. The right-handed photon couples 
to the charm quark loop induced by the 4-quark operator $O_{2} = (\bar
s c)(\bar c b)$, which gives equal rates for $b\to sg\gamma_L$ and  $b\to sg\gamma_R$.
In addition to this leading order effect, 
a significant right-handed photon amplitude in $B\to K^*\gamma$ can appear at subleading 
order in 
$\lqcd/m_b$  from graphs with photon emission from the charm 
quark loop.

The purpose of this paper is to study in more detail the magnitude of the 
leading order effects described above on the time-dependent CP asymmetry in 
$B\to K_S\pi^0\gamma$. In Sec.~II we introduce the effective theory formalism
used in our computation. This is a 
combination of the soft-collinear effective theory (SCET) with the
chiral perturbation theory recently proposed in Ref.~\cite{GPchiral}. 
In Sec.~III the 
helicity amplitudes are written down, and used to compute decay distributions 
for $\bar B\to K_S\pi^0\gamma$ with a right-handed photon. Sec.~IV gives the
results for the time-dependent CP
violation $S_{K_S\pi^0\gamma}$ in the kinematical region with 
an energetic kaon and a soft pion. This has a significant overlap with the region
used in the BELLE and BABAR measurements of the time-dependent CP asymmetry 
into a nonresonant $K_S\pi^0$ state. 
Sec.~V summarizes our results. Readers interested in the phenomenology of the
decay can skip the formalism and proceed directly to Sec.~III.
\vspace{0.5cm}

\section{II. Effective theory formalism}

The exclusive radiative decays $B\to K^{*}\gamma$ can be described
in the large recoil region in factorization. At leading order in 
$\Lambda/m_b$ with $\Lambda \sim 500$ MeV, the existence of such factorization relations has 
been demonstrated in \cite{BeFe,AliP,BeFeSe,BoBu,Sach} at lowest order in $\alpha_s$,
and proven to all orders in $\alpha_s$ using the soft-collinear effective theory
\cite{bpsff,Beneke:2003pa,Hill:2004if,CK,BHN}.

The $b\to s\gamma$ transitions with an energetic $s$ quark
are mediated in SCET$_{\rm I}$ by the effective Lagrangian
\begin{eqnarray}\label{HeffSCET}
H_{\rm eff} &=& N_0
  \Big[ m_b c(\omega)\, \bar s_{n,\omega} \Aslash^\perp P_L\, b_v \\
&+&  b_{1L}(\omega_i)\, O^{(1L)}(\omega_i) 
  + b_{1R}(\omega_i)\, O^{(1R)}(\omega_i) + {\cal O}(\lambda^2) \Big] , \nn
\end{eqnarray}
with $N_0 = \frac{G_F V_{tb} V_{ts}^*\, e}{\sqrt2\, \pi^2} E_\gamma$.
The relevant modes are soft quarks and gluons with momenta $k_s \sim \Lambda$ and 
collinear quarks and gluons along $n$~\cite{CK}. 
We use everywhere the SCET notations in Ref.~\cite{ps1}.
We choose the photon momentum to move along the $-\vec e_3$ direction
$q_\mu = E_\gamma \bar n_\mu$, such that the hadronic system has a large
momentum along the opposite direction $n_\mu$. The hard scale in this problem
is $Q\equiv \bar n\cdot p_{X_s} \sim m_b$, and the expansion parameter in
SCET is $\lambda^2 \sim \Lambda/Q$.
We denoted $n_\mu,\bar n_\mu$ unit light-cone vectors satisfying
$n^2 = \bar n^2 = 0, n\cdot \bar n = 2$. The transverse photon field is 
${\cal A}_\mu^\perp$, and its polarization vectors are $\varepsilon_+ = (0,\frac{1}{\sqrt2},
-\frac{i}{\sqrt2}, 0)$ (right-handed photon) and 
$\varepsilon_- = (0,\frac{1}{\sqrt2}, \frac{i}{\sqrt2}, 0)$ (left-handed photon).

We neglect here and in the following $s$ quark mass effects, which 
can be included straightforwardly \cite{mass}. 
The photon coupling to the spectator quark in the B introduces new
factorizable operators containing 
collinear modes along the photon momentum \cite{GPrad,CK,BHN}. 
These spectator effects do not contribute to the right-handed photon
amplitude, and we return to them below (see the discussion around Eq.~(\ref{HeffIII})).

The first operator in Eq.~(\ref{HeffSCET}) scales like $O(\lambda^0)$ and couples 
only to left-handed photons. The $O(\lambda)$ operators $O^{(1L,R)}$ couple 
to left- and right-handed photons, respectively, and are defined as
\begin{eqnarray}
O^{(1L)}(\omega_1,\omega_2) &=& \bar s_{n,\omega_1}\, \Aslash^\perp 
  \Big[\frac1{\bar n\cdot {\cal P}} ig \Bslash^\perp_n\Big]_{\omega_2}
  P_R\, b_v , \nn\\*
O^{(1R)}(\omega_1,\omega_2) &=& \bar s_{n,\omega_1}
  \Big[\frac1{\bar n\cdot {\cal P}} ig \Bslash^\perp_n\Big]_{\omega_2}
  \Aslash^\perp  P_R\, b_v .
\end{eqnarray}
At lowest order in matching, the Wilson coefficients can be extracted 
from the computations of \cite{3point,AliP,BeFeSe,BoBu,Sach} and are given by \cite{CK}
\begin{eqnarray}\label{b1LR}
c(\omega) &=& C_7^{\rm eff} + O(\alpha_s(m_b))\\
b_{1L}(\omega_1,\omega_2) &=& C_7^{\rm eff} + \frac{2C_2}{3}\kappa
\Big(\frac{-2E_\gamma \omega_2}{m_c^2}\Big) \\
&& + {\cal O}[C_{3-6,8}, \alpha_s(m_b)]\,, \nn\\
b_{1R}(\omega_1,\omega_2) &=& -\, \frac{2C_2}{3}\kappa
\Big(\frac{-2E_\gamma \omega_2}{m_c^2}\Big)
 + {\cal O}[C_{3-6,8}, \alpha_s(m_b)]\,. \nn
\end{eqnarray}
with  
\begin{eqnarray}
C_7^{\rm eff} = C_7 - \frac49 C_3 - \frac43 C_4 + \frac{1}{9} C_5 + \frac{1}{3}C_6
\end{eqnarray}
in the operator basis of Ref.~\cite{BBL}. Beyond tree level the 
Wilson coefficients $c,b_{1L}$ and $b_{1R}$ receive hard corrections $\sim \alpha_s(m_b)$
from charm loops \cite{hard} and from matching onto SCET$_{\rm I}$. The 
$O(\alpha_s(m_b))$ matching corrections are known only for $c(\omega)$ \cite{bfps}, so
for consistency we do not include them in any of the coefficients.
The function $\kappa(x)$ appears in the 3-point function with a charm
loop, and is given by \cite{3point}
\begin{eqnarray}
\kappa(x) =  
\left\{
\begin{array}{ll}
\frac12 - \frac{2}{x} \arctan^2[\sqrt{\frac{x}{4-x}}] & x < 4 \\
\frac12 + \frac{2}{x}\Big( 
\log \left( \frac{\sqrt{x}+\sqrt{x-4}}{2}\right) - \frac{\pi i}{2} \Big)^2 & x > 4 \\
\end{array}
\right.
\end{eqnarray}
In the Wilson coefficients of the $O(\lambda)$ operators
$b_{1L}$ and $b_{1R}$ we neglect small contributions from the penguin operators
$O_{3-6}$ and the gluon dipole operator $O_8$.

After matching onto SCET$_{\rm II}$ \cite{bpsff}, the $b\to s\gamma$ 
effective Lagrangian  (\ref{HeffSCET}) contains both factorizable and 
nonfactorizable operators
\begin{eqnarray}
H_{\rm eff} \to (O_\mu^{\rm nf} + O_\mu^{\rm fact}) {\cal A}^\mu + \cdots
\end{eqnarray}
where the ellipses stand for higher dimension operators. 
The details of
this matching are given in Refs.~\cite{bpsff,ps1,Beneke:2003pa}, and we
give here only the points essential in the following. 

The nonfactorizable operators couple only to the left-handed photon
field, $\varepsilon_+^{*\mu} O^{\rm nf}_\mu = 0$, while the factorizable 
operators couple to both left- and right-handed photons.
Working at tree level in matching SCET$_{\rm I} \to$ SCET$_{\rm II}$, the
factorizable operators are  
\begin{eqnarray}\label{HeffII}
&& O^{\rm fact}_\mu = N_0 \Big[ -\frac{1}{2\omega} 
\int_0^1 \mbox{d}z \mbox{d}x \mbox{d}k_+ b_{1L}(z) J_\perp(x,z,k_+) \nn \\
&&\qquad\qquad \times (\bar q_{k_+} \nslash \gamma^\perp_\mu \gamma_\perp^\lambda P_R b_v)
(\bar s_{n,\omega_1} 
\frac{\bnslash}{2} \gamma^\perp_\lambda q_{n,\omega_2}) \nn\\
&&
- \frac{1}{2\omega}
\int_0^1 \mbox{d} z \mbox{d} x \mbox{d} k_+ b_{1R}(z) J_\parallel(x,z,k_+) \nn \\
&&\qquad\qquad \times (\bar q_{k_+} \nslash \gamma^\perp_\mu P_R b_v)(\bar s_{n,\omega_1} 
\frac{\bnslash}{2} P_L q_{n,\omega_2}) \Big] 
\end{eqnarray}
where we used a momentum space notation for the soft nonlocal operators
\begin{eqnarray}
\bar q^i_{k_+}  b^j_v = \int \frac{d\lambda}{4\pi}
e^{-\frac{i}{2} \lambda k_+} \bar q^i(\lambda n/2) Y_n(\lambda,0) b^j_v(0)\,.
\end{eqnarray}
The functions $b_{1L}(z)$ and $b_{1R}(z)$ appearing here are related to the 
Wilson coefficients
in Eq.~(\ref{HeffSCET}) as $b_i(z) = b_i((1-z)\omega, z\omega)$. The momentum
labels of the collinear fields are parameterized as $\omega_1 = x\omega, 
\omega_2 = -(1-x)\omega$.
We denoted here with $J_{\perp,\parallel}$ jet functions defined as in Ref.~\cite{bprs}.
They have perturbative
expansions in $\alpha_s(\mu_c)$ with $\mu_c^2 \sim Q\Lambda$. At leading order 
they are given by
\begin{eqnarray}\label{jets}
J_\parallel(x,z,k_+)\! = \! J_\perp(x,z,k_+)\! =\! 
\frac{\pi\alpha_s(\mu_c) C_F}{N_c} \frac{1}{\bar x k_+} \delta(x-z)
\end{eqnarray}
The $O(\alpha_s^2(\mu_c))$ corrections to the jet functions have been recently 
obtained in Ref.~\cite{Hill:2004if}.

Another class of factorizable operators not present in Eq.~(\ref{HeffII}) 
arise from the photon coupling to the spectator quarks \cite{GPrad,CK,BHN}. 
(Photon coupling to the final state quarks leads to power suppressed
operators \cite{HLPW}.)
After matching onto SCET$_{\rm II}$, they are 
given at leading order in $\alpha_s(m_b)$ by
\begin{eqnarray}\label{HeffIII}
&& O_\mu^{\rm sp} = \frac{G_F}{2\sqrt2} e \sum_{q=u,d,s} b_{\rm sp}(\omega_i) \\
&& \times \int dk_- J_{\rm sp}(k_-) e_q (\bar q_{k_-} \gamma_\mu
\bnslash \nslash P_L b_v) (\bar s_{n,\omega_1} \frac{\bnslash}{2} P_L q_{n,\omega_2}) \nn
\end{eqnarray}
with $b_{\rm spec}(\omega_i) = V_{ub} V_{us}^* (C_1+C_2/N_c) \delta_{qu} -
V_{tb} V_{ts}^* (C_4+C_3/N_c) + O(\alpha_s(Q))$ and
\begin{eqnarray} 
J_{\rm sp}(k_-) = \frac{1}{k_- + i\epsilon}\Big(1 + \frac{\alpha_s C_F}{4\pi}
(L^2 - 1 
- \frac{\pi^2}{6}) \Big)
\end{eqnarray}
with $L = \log[(-2E_\gamma k_- - i\epsilon)/\mu^2]$
a jet function known to $O(\alpha_s(\mu_c))$ \cite{LPW}. 
The spectator operator couples only to left-handed photons.
For consistency with the other factorizable operators included, we work to
$O(\alpha_s^0(Q))$, but keep terms of $O(\alpha_s(\mu_c))$ in the factorized
amplitude. The dominant term $\sim C_{1,2}$ contributes only to
$\bar B^0 \to K^- \pi^+\gamma$, but not to $\bar B^0 \to K_S \pi^0\gamma$.
In Eqs.~(\ref{b1LR}) we neglected the contributions from $O_{3-6,8}$
to the Wilson coefficients $c(\omega), b_{1L,1R}(\omega_i)$, so for
consistency we neglect such terms also in  Eq.~(\ref{HeffIII}).

The SCET formalism introduced above has been applied to prove factorization
relations for exclusive
semileptonic $B\to M \ell\nu$ and radiative $B\to M \gamma, M \ell^+\ell^-$
decays into one energetic light hadron, with $M$ a light pseudoscalar or 
vector meson \cite{bpsff,ps1,Beneke:2003pa,Hill:2004if,CK,pol,BHN}. In all these cases, 
the transition matrix element is written
as the sum of a soft (nonfactorizable) and hard-scattering (factorizable) terms,
as follows.

The matrix elements of the nonfactorizable operators are parameterized in terms
of soft form factors. In our calculation we require only the matrix element
\begin{eqnarray}\label{Hnf}
\hspace{-0.5cm} \langle K^*(p,\eta) |\bar s_n \varepsslash_-^* P_L b_v |\bar B(v)\rangle \!=\!
(\varepsilon_-^*\cdot \eta^*) \bar n \!\cdot\! p_{K^*} \zeta_\perp^{BK^*} 
\end{eqnarray}
where we use the SCET$_I$ notation for the operators obtained from them
by matching onto SCET$_{II}$. 

The matrix elements of the factorizable operators $O_\mu^{\rm fact}$ 
given in Eq.~(\ref{HeffII}) are given
by convolutions of soft and collinear matrix elements with the Wilson 
coefficients. The matrix elements
are parameterized in terms of light-cone wave functions of the $B$ and
$K^{(*)}$ mesons \cite{CK}. In the $\bar B\to \bar K^*$ transition, 
only the left-handed
photon amplitude is nonvanishing \cite{CK,pol}, and is given by
\begin{eqnarray}\label{Hfact}
H_{-}^{\rm fact}(\bar B\to \bar K^*\gamma_{L}) &\!=\!&
\langle \bar K^*\gamma_{L} |{\cal H}_{\rm eff} |\bar B\rangle \\
&\! =\!&
N_0 m_b m_B \int_0^1 \mbox{d}z b_{1L}(z) \zeta_{J\perp}^{BK^*}(z)\nn
\end{eqnarray}
where the factorizable function $\zeta_{J\perp}^{BK^*}(z)$ is defined as
\begin{eqnarray}\label{zetaJdef}
\zeta_{J\perp}^{BK^*}(z)\! =\! \frac{f_B f_{K^*}^\perp}{m_B}
\int dx dk_+ J_\perp(x,z,k_+) \phi_+^B(k_+) \phi_{K^*}^\perp(x)\nn\\
\end{eqnarray}

In Ref.~\cite{GPchiral} it was proposed to extend the application of
this formalism also to multibody B decays to final states containing
one energetic meson and one soft hadron. 
We summarize briefly the main points of this
approach, before proceeding with the details of the computation.

The matrix element of the nonfactorizable operator is parameterized in a
manner similar to Eq.~(\ref{Hnf}) in terms of a new soft function
\begin{eqnarray}\label{Hnfkpi}
H_{-}^{\rm nf}(\bar B\to K_n\pi\gamma_L) \!=\!
N_0 m_b c(\bar n\!\cdot\! p_{K^*}) \zeta_\perp^{BK\pi}(E_K, E_\pi)
\end{eqnarray}
This nonfactorizable amplitude
couples only to left-handed photons, just as in the
case of the $B\to K^*\gamma$ transition. Furthermore, the soft function
$\zeta_\perp^{BK\pi}$ is related by a symmetry relation to a similar
function appearing in multibody semileptonic decay $B\to \pi\pi \ell\bar\nu$ \cite{GPchiral},
analogous to the appearance of a common nonfactorizable amplitude $\zeta^{BM}$ in
both rare and semileptonic form factors \cite{ffrel,BeFe,bpsff}.
 
The matrix elements of the factorizable operators in Eq.~(\ref{HeffII})
are also given by convolutions of hard, jet and soft factors, as in the case of 
the $B\to K^*$ transition discussed above.
At leading order in $\Lambda/Q$, soft and 
collinear modes decouple in the SCET Lagrangian \cite{bpsfact}, which is the statement of 
soft-collinear factorization. This fact has two important implications.
First, the soft pion does not couple to the collinear meson in the
final state $M$.
Second, the matrix elements of factorizable operators corresponding
to the transition $B\to M_n \pi$ factor as
\begin{eqnarray}\label{factschem}
\langle K_n \pi|O_{\rm fact}|\bar B\rangle = 
\langle K_n |O_{\rm C}|0\rangle \times
\langle \pi |O_S|\bar B\rangle\,,
\end{eqnarray}
and are calculable in terms of the kaon light-cone wave functions and a 
new soft matrix element of the $O_S$ operator in the $B\to \pi$ transition.

The soft operator $O_S$ required here appears in the $b_{1R}$ factorizable
operator. We define it as 
\begin{eqnarray}
O_S(k_+) \! =\! \int \frac{d\lambda}{4\pi} e^{-\frac{i}{2} \lambda k_+}
\bar q(\lambda \frac{n}{2}) Y_n(\lambda \frac{n}{2}, 0) \varepsslash_+ \frac{\nslash}{2}
P_R b_v(0)
\end{eqnarray}
Its $B\to \pi$ matrix element is parameterized in terms of one soft function
$S(k_+, t^2, \zeta)$, defined as
\begin{eqnarray}\label{Sdef}
\langle \pi (p_\pi) |O_S(k_+)|\bar B(v)\rangle = - 2 (\varepsilon_+ \cdot p_\pi)
S(k_+, t^2, \zeta)
\end{eqnarray}
with $t = m_B v - p_\pi$ and $\zeta = n\cdot p_\pi/(n\cdot v)$. The support 
of this function is $-n\cdot p_\pi \leq k_+ \leq \infty$. This is the analog 
for B physics of the generalized parton distributions (GPD), commonly encountered
in nucleon physics.

The complete factorization relation 
for the right-handed photon amplitude can be now written down as
\begin{eqnarray}
&& H_+(\bar B\to \bar K\pi\gamma_R) = 2N_0 f_K (\varepsilon_+ \cdot p_\pi) 
\int_0^1 \mbox{d} z \mbox{d} x b_{1R}(z) \nonumber \\
&& \times
\int_{-p_\pi^+}^\infty \mbox{d} k_+
 J_\parallel(x,z,k_+) S(k_+, t^2, \zeta) \phi_K(x) \,,
\end{eqnarray}

The predictive power of such relations depends on the existence of
reliable information about the soft function $S$.
Eventually the function $S$ should be extracted using B decays data,
or constrained by lattice QCD computations. In the soft pion region,
the soft function $S$ is fixed by chiral symmetry in terms of one of 
the B meson light cone wave functions \cite{GPchiral}.
We will use in this paper the result for $S$ predicted at leading
order in chiral perturbation theory.

The predictions of chiral symmetry for the couplings of Goldstone
bosons are most conveniently derived using chiral perturbation
theory. The applicability of this approach is limited to problems
describing only soft hadrons. The extension to heavy hadrons is
possible, provided that the large scale $m_b$ is eliminated by going
over to HQET. The corresponding chiral effective theory is the
heavy hadron chiral perturbation theory
(HHChPT), and its degrees of freedom are heavy meson spin doublets $H=(B,B^*)$ 
and the Goldstone bosons \cite{wise,BuDo,TM}.

The effective Lagrangian that describes the strong interactions
of the Goldstone bosons with the ground state heavy mesons
is \cite{wise,BuDo,TM}
\begin{eqnarray}\label{Lag}
{\cal L} &=& \frac{f_\pi^2}{8}
\mbox{Tr } [ \partial^{\mu} \Sigma
\partial_{\mu} \Sigma^\dagger ]
-i \mbox{Tr } [\bar H^{a} v_{\mu} \partial^{\mu} H_a] \\
& &+ \frac{i}{2} \mbox{Tr} [\bar H^{a} H_b] v^{\mu} \left[ \xi^\dagger
\partial_{\mu} \xi + \xi \partial_{\mu} \xi^\dagger \right]_{ab} \nonumber\\
& &+\frac12 ig \mbox{Tr } [\bar H^{a} H_b \gamma_{\nu} \gamma_5]
\left[\xi^\dagger \partial^{\nu} \xi - \xi \partial^{\nu}
\xi^\dagger \right]_{ab} + \cdots \nonumber 
\end{eqnarray}
where the ellipsis denote light quark mass terms, $O(1/m_b)$
operators associated with the breaking of heavy quark spin symmetry,
and terms of higher order in the derivative expansion.
The pseudo-Goldstone bosons appear in the Lagrangian through
$\xi = e^{i\Pi /f_\pi}$ ($\Sigma=\xi^2$) where
\begin{eqnarray}
\Pi = \left(
\begin{array}{ccc}
 {1\over\sqrt2}\pi^0 +
{1\over\sqrt6}\eta &
\pi^+ & K^+ \\
\pi^-& -{1\over\sqrt2}\pi^0 + {1\over\sqrt6}\eta&K^0 \\
K^- &\bar K^0 &- {2\over\sqrt6}\eta \\
\end{array}
\right)
\end{eqnarray}
with the pion decay constant $f_\pi \simeq 135$~MeV. These fields transform as
\begin{eqnarray}
\xi \to L \xi U^\dagger = U \xi R^\dagger
\end{eqnarray}
under chiral $SU(3)_L \times SU(3)_R$ transformations.
The superfield $H_a$ contains the pseudoscalar and vector
heavy meson fields $\bar B_a$ and $\bar B^{*}_{a\mu}$ with 
velocity label $v_\mu$
\begin{eqnarray}
H_a = \frac{1+\vslash }{2} \left[ \bar B^*_{a \mu} \gamma^{\mu}
- \bar B_a \gamma_5 \right]. 
\end{eqnarray}
The flavor index runs over $a=1,2,3$ corresponding to 
$\bar B_a = (B^-, \bar B^0, \bar B_s)$.
Under chiral $SU(3)_L\times SU(3)_R$, the superfield $H_a$ transforms as
\begin{eqnarray}
H_a \rightarrow   H_b \ U^\dagger_{ba} .
\end{eqnarray}
The numerical value of the coupling $g=0.5\pm 0.1$ is taken to cover a range 
compatible with its determination
from $D^*\to D\pi$ decays $g= 0.59 \pm 0.08$~\cite{CLEO} and lattice QCD 
$g = 0.48\pm 0.03 \pm 0.11$ \cite{lqcd1}, $g= 0.42\pm 0.04\pm 0.08$ \cite{lqcdrev}.

We consider next the matrix element $\langle \pi|O_S|\bar B\rangle$ in
chiral perturbation theory. This
requires the chiral representation of the nonlocal soft operators $O_S$. 
Consider light-cone nonlocal heavy-light bilinears of the form
\begin{eqnarray}
O^a_{L,R}(k_+) \!= \!\int \frac{\mbox{d} x_-}{4\pi} \! e^{-\frac{i}{2} k_+ x_-}
\bar q^a(x_-) Y_n(x_-,0) P_{R,L} \Gamma b_v(0)\,.\nonumber\\
\end{eqnarray}
Under the chiral group they transform as
$(\mathbf{\overline{3}}_L, \mathbf{1}_R)$ and
$(\mathbf{1}_L, \mathbf{\overline{3}}_R)$, respectively.
For each case, there is a unique operator in the effective theory
with the correct transformation properties~\cite{GPchiral}
\begin{eqnarray}\label{JOL}
&& O^a_{L}(k_+)  = \frac{i}{4} 
\mbox{Tr } [\hat \alpha_L(k_+) P_R\Gamma H_b \xi^{\dagger }_{ba}],\\
\label{JOR}
&& O^a_{R}(k_+)  = \frac{i}{4} 
\mbox{Tr } [\hat \alpha_R(k_+) P_L\Gamma H_b \xi_{ba}]
\end{eqnarray}
The common matrix $\hat \alpha_L(k_+)=\hat \alpha_R(k_+)=\hat \alpha(k_+)$ 
is given by
\begin{eqnarray}\label{alpha}
\hat\alpha(k_+)  = f_B \sqrt{m_B} [\bnslash \phi_+^B(k_+) + 
\bnslash \phi_-^B(k_+)]
\end{eqnarray}
where $\phi_\pm^B(k_+)$ are the usual light-cone B meson wave functions, 
defined by \cite{BeFe}
\begin{eqnarray}
&& \int \frac{dz_- }{2\pi} e^{-\frac{i}{2} k_+ z_-}
\langle 0 | \bar q_i(z_-)Y_n(z_-,0) b_v^j(0)|\bar B(v)\rangle = \\
&& -\frac{i}{4}f_B m_B \left\{
\frac{1+\vslash}{2} [\bnslash n\cdot v \phi_+^B(k_+) +
\nslash \bn\cdot v \phi_-^B(k_+) ]\gamma_5 \right\}_{ji} \nonumber
\end{eqnarray}
The operators in Eqs.~(\ref{JOL}), (\ref{JOR}) with (\ref{alpha})
can be used to compute the matrix elements of $O_{L,R}$
on states with a $B$ meson and any number of pseudo Goldstone bosons.
In particular, they give the following prediction for the soft function
$S(k_+,t^2,\zeta)$ defined in Eq.~(\ref{Sdef}) at leading order in
the chiral expansion \cite{GPchiral}
\begin{eqnarray}
S(k_+,t^2,\zeta) = \frac{g f_B m_B}{f_\pi} \frac{1}{v\cdot p_\pi + \Delta} \phi_+^B(k_+)
\end{eqnarray}
We will use this result in the numerical evaluations of this paper.
No information can be obtained using chiral symmetry about the $S$
function for $-n\cdot p_\pi < k_+ < 0$. For soft pions this contribution
is likely to be small, so it will be neglected in the following.

Finally, we comment briefly on previous applications \cite{HYC} of HHChPT to B decays into 
multibody final states containing soft Goldstone bosons. 
These applications involve the generalization to $B^*$ decays of the
factorization formula for nonleptonic B decays,
with the $B^*$ appearing in intermediate states of pole diagrams. 
The usual HHChPT \cite{wise,BuDo,TM} methods are applied to compute the
pion coupling in both pole diagrams and in contact diagrams.

There are several issues with such a simplified approach: i) the application of chiral
perturbation theory to the nonfactorizable operators $O_{\rm nf}$ 
contributing to the $B\to \pi$ transition with an energetic pion is 
problematic. Since these operators couple to both soft and
collinear modes, loop corrections to their matrix elements do not have a 
well-behaved power counting.
In our approach these contributions are simply parameterized by new
soft functions $\zeta_\perp^{BK\pi}$, which are related by symmetry relations
to similar matrix elements appearing in other processes.
ii) the pion contact terms, such as that in Fig.~2b, can be computed only
for the factorizable operators, (but not for the entire weak vertex),
and are given by factorization relations.

In the next section we will combine the pieces of the factorizable
amplitudes, add in the nonfactorizable amplitude and write down the complete 
result for the multibody $B\to K\pi\gamma$ amplitudes.
\vspace{0.5cm}

\section{III. Helicity amplitudes and decay rates}

\begin{figure}
\centering
\begin{picture}(300,170)
\put(100,55){\makebox(50,50){\epsfig{figure=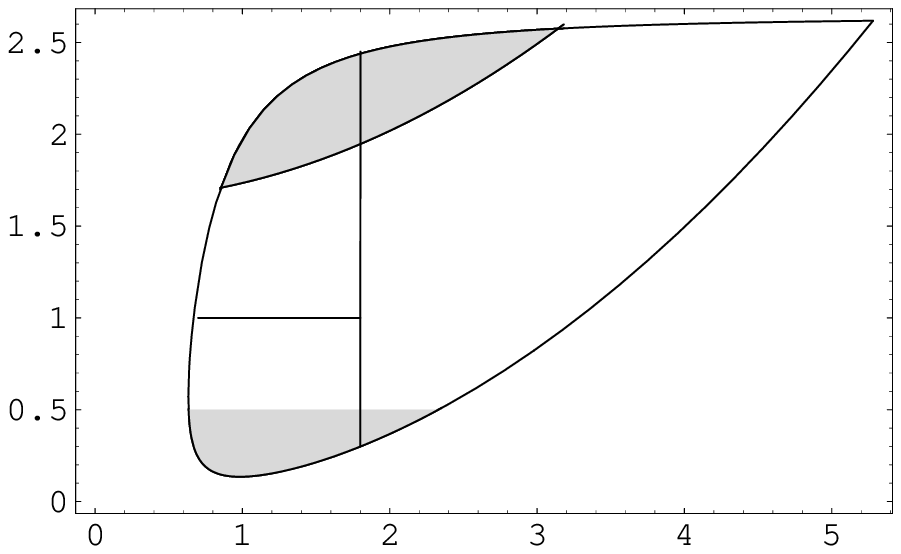,height=5cm}}}
\put(-5,125){
{\large 
$E_\pi$
}}
\put(-15,105){
{\large 
(GeV)
}}
\put(80,-5){
{\large 
$M_{K\pi}\mbox{ (GeV)}$
}}
\put(72,55){
{\large 
$\mbox{ I}$
}}
\put(70,80){
{\large 
$\mbox{ II}$
}}
\put(78,118){
{\large 
$\mbox{ III}$
}}
\end{picture}
\caption{\label{fig2} 
The phase space of the $\bar B^0 \to K_S \pi^0\gamma$ decay in
variables $(M_{K\pi}, E_\pi)$. The vertical line denotes the maximum
$K\pi$ invariant mass used in the BABAR and BELLE measurements.
The 3 regions shown correspond to (I) soft pion $E_\pi \sim \Lambda$;
the shaded region $E_\pi\leq 0.5$ GeV 
(implying $E_K > 2.18$ GeV) shows the region of applicability of ChPT; 
(II) collinear pion and kaon $E_\pi \sim Q,
E_K > 1$ GeV; (III) soft kaon $E_K < 1$ GeV (implying $E_\pi > 1.7$ GeV).
}
\end{figure}

We will use the formalism described in Sec.~II to compute the amplitude
for the decays $\bar B\to K_S \pi^0 \gamma$ and $\bar B\to K^- \pi^+ \gamma$
in the kinematical region with one
energetic (collinear) kaon and one soft pion.
To establish the region of validity of our computation, 
we show in Fig.~1 the phase space for this decay, in 
variables $(M_{K\pi}, E_\pi)$, with $M_{K\pi}^2 = (p_K + p_\pi)^2$.

We distinguish three distinct regions for the pion and kaon energies in
$\bar B\to K\pi\gamma$ decay (see Fig.~1):

I) $E_\pi \sim \Lambda, E_K \sim Q$

II) $E_\pi \sim Q, E_K \sim Q$

III) $E_\pi \sim Q, E_K \sim \Lambda$

These three regions are treated differently in the SCET,
and the heavy quark mass scaling of the decay amplitudes
is correspondingly different in each of them, as follows.

The region (I) contains a soft pion and an energetic kaon.
Part of this region, but not all, can be treated using 
the SCET + ChPT combination considered in this paper. We subdivide it
into the soft pion region with $E_\pi < 0.5$ GeV, where chiral
perturbation theory is valid, and the
intermediate pion region $0.5 \mbox{GeV } < E_\pi < 1.0-1.5$ GeV
(which we will call, for lack of a better name, the hard-soft pion region).

The region (II) includes collinear pion and kaons. In general
this configuration can have the kaon and pion momenta moving in different
directions forming a large angle in the B rest frame $\theta_{\pi K} \sim
O(1)$. In our case, the experimental constraint
$M_{K\pi} < 1.8$ GeV forces the angle to be small $\theta_{\pi K} \sim 
O(\Lambda/Q)$ (valid for $M_{K\pi}^2 \sim \Lambda Q$), such that the $\pi, K$
constituent partons can be described by collinear fields with a common $n$.

The region (III) contains an energetic pion and a soft kaon $E_K \sim 
\Lambda, E_\pi \sim Q$. This region is not described by the
leading order SCET operators in Sec.~II, and the corresponding amplitudes
are suppressed by at least one power of $\Lambda/Q$ relative to those in
regions (I) and (II).

We start by writing down the helicity amplitudes for the $B\to K^*\gamma$ decay
at leading order in $\lqcd/m_b$, by combining the partial results in Sec.~2.
At this order, only the left-handed photon amplitude is nonvanishing 
\begin{eqnarray}\label{HRleading}
&& H_+(\bar B\to \bar K^*\gamma_R) = 0\\
\label{Hminus}
&& H_-(\bar B\to \bar K^*\gamma_L) = N_0 m_b m_B  \\
&& \qquad \times
\Big( c(m_B) \zeta_\perp^{BK^*} + \int_0^1 \mbox{d}z b_{1L}(z) \zeta_{J\perp}^{BK^*}(z)
\Big)\nn \\
&& \hspace{2.6cm} \equiv N_0 m_b m_B C_7^{\rm eff} g_+^{\rm eff}(0)\nn\,.
\end{eqnarray}
We defined here the effective tensor form factor $g_+^{\rm eff}(0)$, which 
absorbs the contributions of the operators other than $O_7$. Similar 
factorization relations
are expected to hold also for the $\bar B\to K\pi \gamma$ transition in region (II),
with the $K^*$ light-cone wave function replaced by a two-body $K\pi$ light-cone wave
function. We do not pursue this further here, but note only that the vanishing of the
right-handed photon amplitude at LO observed in Eq.~(\ref{HRleading}) should hold also 
for $B\to K_n \pi_n$ in the region (II). Technically, this follows from the vanishing 
of the $\bar B\to 0$ matrix 
element of the soft operator in Eq.~(\ref{HeffII}) multiplying $b_{1R}$.

The amplitudes for the multibody transition
$\bar B\to \bar K\pi\gamma$ in region (I) are given at leading order in $\Lambda/m_b$
and in the chiral expansion by the graphs in Fig.~2
\begin{eqnarray}\label{Hplus}
\mbox{(I)} &:& H_+(\bar B\to \bar K\pi\gamma_R) = N_0 \frac12 m_B^2 S_R(p_\pi)\\
&& \qquad \times  \int_0^1 dz
b_{1R}(z) \zeta_{J\parallel}^{BK}(z)\nn\\
&& H_-(\bar B\to \bar K\pi\gamma_L) = 
N_0 \bar n\cdot p_K c(m_B) \zeta_\perp^{BK\pi}(E_K,E_\pi) \nonumber \\
&& \label{Hpm}
\end{eqnarray}
The nonfactorizable operators $O_{\rm nf}^\mu$ contribute  only to the
left-handed photon amplitude, and the corresponding matrix element
is parameterized by $\zeta_\perp^{BK\pi}$. On the other hand, the
factorizable operators $O_{\rm fact}^\mu$ contribute only to the right-handed
photon amplitude.
(We consider only $\bar B^0 \to K_S\pi^0\gamma$ decays, for which the spectator 
factorizable operators $O_{\rm spec}$ do not contribute in the approximation used
here of neglecting $O_{3-6}$.)

Note the appearance of a nonvanishing right-handed photon amplitude at 
leading order in the region (I). This amplitude is factorizable and can be 
computed as explained in Sec.~II. The HHChPT diagrams required
for its computation are shown in Fig.~2a.
The factorizable function $\zeta_{J\parallel}^{BK}(z)$ 
appearing here is defined in analogy with the
function in Eq.~(\ref{zetaJdef}), with the replacements 
$f_{K^*}^\perp \to f_{K}\,,
\phi^\perp_{K^*}(x)\to \phi_{K}(x)\,, 
J_\perp \to J_\parallel $.
The dependence on the pion momentum is contained in the soft functions 
$S_{R}(p_\pi)$ given by
\begin{eqnarray}
S_R(p_\pi) &=&  \frac{g}{f_\pi} 
\frac{\varepsilon_+ \cdot p_\pi }{E_\pi + \Delta} 
\end{eqnarray}
with $\Delta = m_{B^*} - m_B = 50$ MeV.

\begin{figure}
\includegraphics[height=1.5cm]{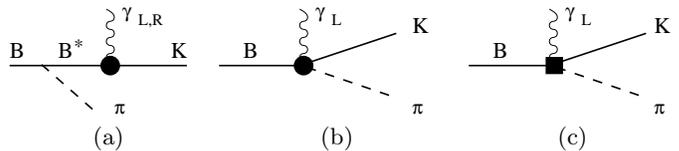}\\
\hspace{0.1cm} (a)\hspace{2.5cm} (b) \hspace{2.5cm} (c) 
\caption{\label{fig1} Diagrams showing leading order contributions to the decay 
$\bar B \to K \pi\gamma$ with one collinear kaon and a soft pion. The 
filled circle in (a) and (b) represents a factorizable operator $O_{\rm fact}$,
while the filled square in (c) represents a nonfactorizable operator $O_{\rm nf}$.
}
\end{figure}

Finally, in the kinematical region (III) with one soft kaon and an energetic
pion, the effective Lagrangian Eq.~(\ref{HeffSCET}) does not apply. The 
leading SCET$_{\rm I}$ operator mediating such a transition contains
the $\bar s\Gamma b_v$ soft current, with 
at least two insertions of the soft-collinear Lagrangian, acting on the
spectator quark
\begin{eqnarray}
T\{ (\bar s\Gamma b_v)\,, i{\cal L}^{(1)}_{q\xi}\,, i{\cal L}^{(1)}_{q\xi} \}
\end{eqnarray}
This is suppressed by at least $\lambda^2 \sim \Lambda/Q$ relative
to the operators in Eq.~(\ref{HeffSCET}), which implies that the
decay amplitudes in this region must be power suppressed relative to those
in regions (I) and (II).

We summarize the different contributions enumerated above by showing
them in graphical form in Fig.~2. We emphasize our different treatment of 
the factorizable and nonfactorizable
operators: the matrix elements of the factorizable operators
are computed in chiral perturbation theory, and include the $B^*$ pole
and contact terms (Fig.~2(a), (b)). The matrix element of the nonfactorizable
operator (Fig.~2(c)) is parameterized in terms of a new soft function $\zeta_\perp^{BK\pi}$.
This new soft function appears only in the left-handed photon amplitude, and 
is related by symmetry relations to a similar soft function which can be determined
in principle from $\bar B\to \pi_n \pi \ell \bar\nu$ \cite{GPchiral}.

The only region where the right-handed photon amplitude 
$\bar B^0\to K_S\pi^0 \gamma$ contributes
at leading order in $\Lambda/Q$ is the region (I) with a soft pion. 
In region (II) the right-handed
amplitude is suppressed by $\Lambda/Q$ relative to the left-handed amplitude,
and in region (III) both amplitudes are suppressed by at least $\Lambda/Q$.
The contribution of the region (I) with a hard-soft pion 0.5 GeV  $< E_\pi < 1.0-1.5$ GeV
will be estimated by assuming the validity of HHChPT in the entire region (I).
We proceed to compute the right-handed photon effect in region (I) on the
decay rates and time-dependent CP asymmetry.

The left-handed photon amplitude
$H_-(\bar B\to \bar K\pi\gamma_L)$ in Eq.~(\ref{Hpm}) (region (I)) does not include the
$K^*$ pole contribution, although this likely dominates numerically
in the resonant region $M_{K\pi} \sim M_{K^*}$. This
contribution is parametrically suppressed by $\lqcd/m_b$ in the soft pion
kinematical region. The reason for this is that by soft-collinear
factorization at leading order in $\lqcd/m_b$, the coupling of a
soft pion to two collinear hadrons,  $K^*_n K_n\pi_S$, must be power suppressed.
On the other hand, the $K^*$ pole contribution is numerically
enhanced by the $K^*$ propagator, so we will include it 
in our computation, despite being formally of higher order in $\lqcd/m_b$ 
relative to the latter.

In the absence of data on $\zeta_\perp^{BK\pi}$, we will model it 
by a $K^*$ pole contribution.
We introduce the following model for the 
$\bar B^0\to \bar K\pi\gamma$ decay  amplitudes in the
kinematical region (I) with one soft pion and an energetic kaon 
(this is similar to the model used in Ref.~\cite{fba}). In the left-handed
photon amplitude we neglect the nonresonant contribution and keep 
only the $K^*$ pole term
(Fig.~2c)
\begin{eqnarray}\label{modelL}
&& H^{\rm model}_-(\bar B\to K^-\pi^+\gamma_L) =\\
&& \hspace{2cm} H_-^{BK^*} g_{K^* K\pi}
(\varepsilon_-\!\cdot\! p_\pi) \mbox{BW}_{K^*}(M_{K\pi}) \nonumber
\end{eqnarray}
with $H_-^{BK^*} = A(\bar B\to K^*\gamma_L)$ the 1-body helicity amplitude given 
in Eq.~(\ref{Hminus}), and $\mbox{BW}_{K^*}(M_{K\pi}) = 
[M_{K\pi}^2 - M_{K^*}^2 + i M_{K^*} \Gamma_{K^*} ]^{-1}$ the Breit-Wigner 
function for a $K^*$ resonance. The amplitude for $\bar B^0 \to K_S\pi^0\gamma_L$
is given by Eq.~(\ref{modelL}) multiplied by $1/2$.
The $K^{*0}K^+\pi^-$ coupling with a charged pion can be extracted from the 
total $K^*\to K\pi$
decay width, $\Gamma = g_{K^*K\pi}^2 p_\pi^3/(16 \pi m_{K^*}^2)$, which gives
$g_{K^*K\pi} = 9.1$. 

\begin{figure}
\centering
\begin{picture}(300,170)
\put(100,55){\makebox(50,50){\epsfig{figure=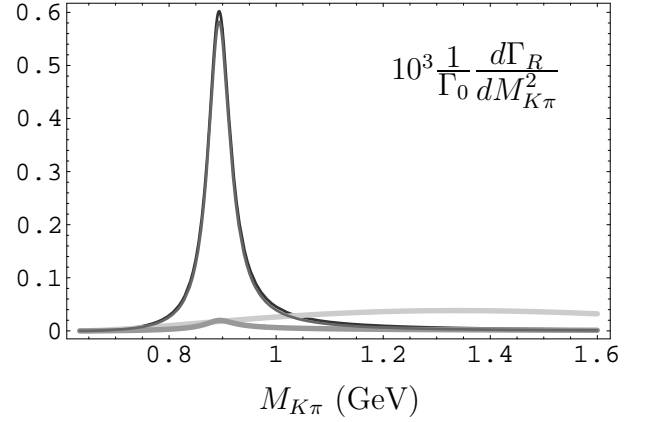,height=5cm}}}
\put(150,120){
{\large 
$10^3 \frac{\displaystyle 1}{\displaystyle \Gamma_0}
\frac{\displaystyle d\Gamma_R}{\displaystyle dM_{K\pi}^2}$
}}
\put(100,-5){
{\large 
$M_{K\pi}\mbox{ (GeV)}$
}}
\end{picture}
\caption{\label{fig3} 
The decay rate ($\times 10^3$) with a right-handed 
photon $\bar B^0 \to K_S \pi^0\gamma_R$, normalized to the $B\to K^* \gamma_L$ rate,
with a cut on the pion energy $E_\pi \leq 0.5$ GeV, and its components, computed as
described in the text.
The black curve gives an upper bound on the total decay rate.
The dark grey curve shows the $B^*-K^*$
interference term, and the light grey curve shows the contribution of the $B^*$ pole graph
(magnified by a factor of $10^3$ relative to the other curves).
}
\end{figure}

The right-handed photon amplitude is given by the sum of the $K^*$ and $B^*$
resonant terms
\begin{eqnarray}\label{modelR}
&& H^{\rm model}_+(\bar B^0\to K^-\pi^+\gamma_R)  \\
&& \hspace{1cm} = N_0  \frac{gm_B^2}{2f_\pi} 
\frac{\varepsilon_+ \cdot p_\pi }{E_\pi + \Delta}
\int_0^1 dz b_{1R}(z) \zeta_J^{BK}(z) \nonumber \\
&& \hspace{1cm} +\, H_+^{BK^*} g_{K^* K\pi}
(\varepsilon_+ \cdot p_\pi) \mbox{BW}_{K^*}(M_{K\pi}) \nn
\end{eqnarray}
The $\bar B^0\to K_S \pi^0 \gamma_R$ amplitude has an additional factor of
$1/2$.
We included here also a nonvanishing resonant $ \bar B \to K^*\gamma_R$
right-handed photon amplitude, which is introduced by a nonvanishing
strange quark mass, and by power suppressed contributions
neglected in Eq.~(\ref{HRleading}). We will parameterize it as
\begin{eqnarray}\label{HRLmodel}
\frac{H_+^{BK^*}}{H_-^{BK^*}} = \frac{m_s}{m_b} + h_s e^{i\phi_s}
\end{eqnarray}
The potentially leading mechanism contributing to $h_s$ 
has been identified in Ref.~\cite{pol}, and it
arises from charm loops coupling to the B and $K^*$ through soft gluons. 
Although no first principles calculation of this parameter is
yet available, it can be estimated from a simple power counting argument as
\begin{eqnarray}
h_s \sim \frac13 \frac{C_2}{C_7} \frac{\Lambda}{m_b} \sim 0.09
\end{eqnarray}
In our numerical evaluation we will use $h_s = 5\%$, keeping in mind
that this estimate is on the lower side of the dimensional estimate.

In the remainder of the paper we will use the model described by 
Eqs.~(\ref{modelL})-(\ref{HRLmodel}) 
to compute distributions and decay rates for the $B\to K_S \pi^0\gamma$ decay.
We start by computing the right-handed photon rate; although not directly 
observable, this quantity will illustrate the relative importance of the different
mechanisms contributing to the wrong-helicity photon amplitude.

The $\bar B\to \bar K\pi\gamma$ decay rate is given by
\begin{eqnarray}\label{rate}
\frac{d^2\Gamma(B\to K\pi\gamma)}{dE_\pi dM_{K\pi}^2}  =
\frac{1}{2(4\pi)^3 m_B^2} (|H_+|^2 + |H_-|^2)
\end{eqnarray}
where $H_\pm$ are given in Eqs.~(\ref{modelL}) and (\ref{modelR}), respectively. 
In the limit of a very narrow $K^*$ the integrations over
$(M_{K\pi}, E_\pi)$ can be performed exactly,
and the well-known result for the $\bar B\to \bar K^*\gamma$ rate is recovered
\begin{eqnarray}
&& \int dE_\pi dM_{K\pi}^2 
\frac{d^2\Gamma(\bar B\to \bar K\pi\gamma_L)}{dE_\pi dM_{K\pi}^2}  =\\
&& \hspace{2cm} \frac{E_\gamma^{(0)}}{8\pi m_B^2} |H_-(\bar B\to K^*\gamma_L)|^2 \equiv 
\Gamma_0 \nn
\end{eqnarray}
Here $E_\gamma^{(0)}=(m_B^2-m_{K^*}^2)/(2m_B)$ denotes the photon energy
corresponding to the 2-body kinematics.

It is convenient to express the $\bar B\to K^-\pi^+\gamma_R$ decay rate
by normalizing it to the $\bar B\to K^*\gamma_L$ decay rate
(a factor of $1/4$ has to be added for $\bar B^0 \to K_S\pi^0\gamma_R$)
\begin{eqnarray}\label{d2G}
&&\frac{1}{\Gamma_0}\frac{d^2\Gamma_R}{dM_{K\pi}^2 dE_\pi} = 
\frac{|\vec p^\perp_\pi\,|^2}{2(4\pi)^2 E_\gamma^{(0)}} \\
&& \times
\left|
\frac{E_\gamma}{E_\gamma^{(0)}}
\frac{g\kappa e^{i\phi}}{2f_\pi (E_\pi + \Delta)}
+
\frac{g_{K^*K\pi}(\frac{m_s}{m_b} + h_s e^{i\phi_s})}
{M_{K\pi}^2-M_{K^*}^2 + i M_{K^*}\Gamma_{K^*} } \right|^2 \nn
\end{eqnarray}
with $M_{K\pi}^2 = m_B^2 - 2m_B E_\gamma$. The two terms give the 
contributions of the $B^*$ resonant pole, and that of the $K^*$ 
resonant right-handed amplitude.
The hadronic dynamics in the $B^*$ resonant contribution
enters through the RG invariant ratio
\begin{eqnarray}
\kappa e^{i\phi}  &=& \frac{m_B}{m_b C_7^{\rm eff} g_+^{\rm eff}(0)}
\int_0^1 dz b_{1R}(z) \zeta_J^{BK}(z) \\
&=& -0.013 - 0.045 i \nn
\end{eqnarray}
In the numerical evaluation of this parameter
we used the tree level result for the jet functions Eq.~(\ref{jets})
and the lowest order matching
result for $b_{1R}$ from Eq.~(\ref{b1LR}). The remaining required parameters 
are listed in Table II.

\begin{table}
\caption{\label{table} Input parameters used in the numerical computation,
and results for the effective Wilson coefficients and factorizable matrix 
elements. The values of the
effective Wilson coefficients are quoted at the scale $\mu = 4.8$ GeV.
The strange quark mass is taken from \cite{lqcdrev}.}
\begin{ruledtabular}
\begin{tabular}{cc|cc}
$m_b^{\rm pole}$ & $4.8$ GeV & $C_2$ & $1.107$ \\
$\bar m_c(\bar m_c)$ & $1.4$ GeV  & $C_7$ & $-0.343$ \\
$m_s(2\mbox{ GeV})$ & $78\pm 10$ MeV & $f_B$   & $200$ MeV  \\
$g_+^{\rm eff}(0)$ & $0.3$ & $f_K$   & $170$ MeV  \\
$\langle k_+^{-1}\rangle_B^{-1}$ & $350$ MeV & $g_{K^*K\pi}$ & $9.1$  \\
$g$ & $0.5\pm 0.1$ & $\kappa e^{i\phi}$  & $-0.013 - 0.045 i$  \\
\end{tabular}
\end{ruledtabular}
\end{table}

The result Eq.~(\ref{d2G}) can be used to compute the energy spectrum 
and integrated rate with a right-handed photon, with an upper cut-off on the pion energy.
The interference of the two terms depends sensitively on the unknown strong 
phase $\phi_s$.
For this reason, we will give only an upper bound on this rate, obtained by
assuming that the two terms have the same strong phase and interfere constructively.
The resulting photon energy spectrum and its components are shown in Fig.~3,
for the central values of the parameters.
For completeness, we quote also the fraction of events 
which survive a pion energy cut $n_{\rm cut}(E_\pi^{\rm max})$. This can
be computed from the $K^*$ pole contribution to $H_-(\bar B \to \bar K\pi\gamma)$ 
and is:
$n_{\rm cut}(0.5 \text{~GeV}) = 11.4\% \,, 
n_{\rm cut}(1 \text{~GeV}) = 50.5\%\,, 
n_{\rm cut}(1.5 \text{~GeV})=88.4\%$. 

\section{IV. Time-dependent CP asymmetry}

We compute in this section the
mixing-induced CP violating parameter $S_{K_S\pi^0\gamma}$ in the Standard
Model. 
We start by defining the time-independent amplitudes
\begin{eqnarray}
\bar A_L &=& H(\bar B^0 \to K_S\pi^0 \gamma_L) \\
\bar A_R &=& H(\bar B^0 \to K_S\pi^0 \gamma_R) \\
A_L &=& H(B^0 \to K_S\pi^0 \gamma_L) \\
A_R &=& H(B^0 \to K_S\pi^0 \gamma_R) 
\end{eqnarray}
Since the $b\to s$ transition is CP conserving, there are relations
among these amplitudes, such that only two of them are independent.
We choose the independent amplitudes to be $\bar A_L, \bar A_R$,
and obtain the remaining two amplitudes from them by CP transformations.
Charge conjugation exchanges particles and antiparticles, and
parity takes $\gamma_L \leftrightarrow \gamma_R$ and changes the
directions of momenta. We apply a rotation by $180^\circ$ around the
$x$ axis, which restores the momenta to their original directions.
The effect of the rotation is to multiply the amplitudes with
$\Pi_i \pi_i (-)^{J_i} = +1$ (with $J_i^{\pi_i}$ the spin-parity
of the particles), and exchange $\varepsilon_+ \leftrightarrow \varepsilon_-$.
This gives 
\begin{eqnarray}
A_L = \frac{\varepsilon_+\cdot p_\pi}{\varepsilon_-\cdot p_\pi} \bar A_R \,, \qquad
A_R = \frac{\varepsilon_-\cdot p_\pi}{\varepsilon_+\cdot p_\pi} \bar A_L
\end{eqnarray}

\begin{figure}[b!]
\centering
\begin{picture}(300,170)
\put(110,55){\makebox(50,50){\epsfig{figure=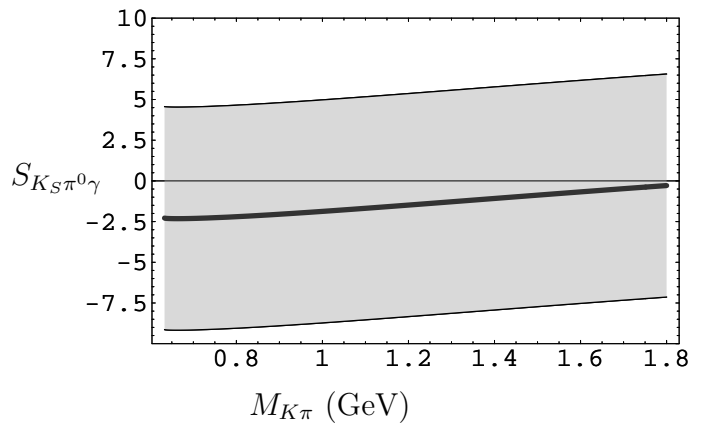,height=5cm}}}
\put(-10,82){
{\large 
$S_{K_S\pi^0\gamma}$
}}
\put(80,-5){
{\large 
$M_{K\pi}\mbox{ (GeV)}$
}}
\end{picture}
\caption{\label{fig5} 
The time-dependent CP asymmetry parameter $S_{K_S\pi^0\gamma}$ (in percent)
as a function of $M_{K\pi}$. The three lines correspond to 
(from bottom to top): $h_s\cos\phi_s = -0.05, 0, 0.05$. We used here 
$E_\pi^{\rm max} = 0.5$ GeV.}
\end{figure}

The time-dependent differential rate has a form
similar to Eq.~(\ref{rate}) (with $i=L,R$)
\begin{eqnarray}\label{rateTD}
&& \frac{d^2\Gamma(B^0(t)\to K_S\pi^0\gamma_i)}{dE_\pi dM_{K\pi}^2}  =
\frac{1}{2(4\pi)^3 m_B^2} (|A_i|^2 + |\bar A_i|^2) \nn \\
&& \qquad \times \frac12 e^{-\bar\Gamma t}
 \left\{ 1 + C_i \cos \Delta mt - S_i \sin \Delta mt \right\} 
\end{eqnarray}
with 
\begin{eqnarray}
&& C_{i}(E_\pi, M_{K\pi}) = \frac{|A_i|^2-|\bar A_i|^2}{|A_i|^2+|\bar A_i|^2}\\
&& S_{i}(E_\pi, M_{K\pi}) =  
2\frac{\mbox{Im } (e^{-2i\beta} \bar A_{i} A^*_{i})}{|A_i|^2+|\bar A_i|^2}
\end{eqnarray}
From this expression, results for the CP violating coefficients
integrated over parts of the phase space can be straightforwardly obtained.

The BELLE and BABAR Collaborations measured the $S$ and $C$ parameters 
integrated over all $E_\pi$ and a range of $M_{K\pi}$. 
We compute the SM values of these parameters, by integrating the
time-dependent distribution Eq.~(\ref{rateTD}) with appropriate cuts.
Applicability of the chiral perturbation theory computation of the
$\bar A_R$ amplitude requires that we restrict the pion energy in the
B rest frame by $E_\pi < E_\pi^{\rm max}$, with $E_\pi^{\rm max} = 500$ MeV.

Integrating over $E_\pi$ with an upper cut-off $E_\pi^{\rm max}$, and summing
over the photon polarizations gives the mixing-induced CP asymmetry parameter
$S_{K_S\pi^0\gamma}(M_{K\pi})$ 
\begin{eqnarray}
&& S_{K_S\pi^0\gamma}(M_{K\pi}) = - 2\sin 2\beta \Big\{ \frac{m_s}{m_b} + h_s 
\cos\phi_s \\
&&\! + \! \frac{gI(M_{K\pi})}{2f_\pi g_{K^*K\pi}} \frac{E_\gamma}{E^{(0)}_\gamma}
[(M^2_{K\pi} - M_{K^*}^2) \mbox{Re } \kappa\! -\! 
M_{K^*}\Gamma_{K^*} \mbox{Im } \kappa ] \Big\}
\nonumber
\end{eqnarray}
The first two terms represent the resonant $B\to K^*\gamma$
effect, and the last term is the nonresonant contribution.
The dependence on the pion energy cut-off is contained in
\begin{eqnarray}
I(M_{K\pi}, E_\pi^{\rm max}) =  
\frac{\int^{E_\pi^{\rm max}} dE_\pi \frac{|\vec p_\pi^{\perp}|^2}{E_\pi+\Delta}}
{\int^{E_\pi^{\rm max}} dE_\pi |\vec p_\pi^{\perp}|^2}
\end{eqnarray}

We show in Fig.~4 results for the $S_{K_S\pi^0\gamma}$
parameter as a function of $M_{K\pi}$, integrated with an upper
pion energy cutoff
$E_\pi^{\rm max} = 0.5$ GeV. We used in this computation the central value
for $\sin 2\beta = 0.685 \pm 0.032$ as measured in the charmonium system 
\cite{hfag}. The effect of the nonresonant contribution is to
introduce a mild dependence of the asymmetry on $M_{K\pi}$. 
\vspace{0.5cm}


Finally, we integrate also over $M_{K\pi} =(m_K+m_\pi, 1.8)$ GeV to obtain the
inclusive CP asymmetry parameter (for an upper pion energy cut)
\begin{eqnarray}\label{Stot}
S_{K_S\pi^0\gamma} &=& - 2\sin 2\beta \Big\{ \frac{m_s}{m_b} + h_s 
\cos\phi_s \\
 & & +  \frac{g}{2f_\pi g_{K^*K\pi}} 
\mbox{ Re }[\kappa I_2(E_\pi^{\rm max})] \Big\}
\nonumber
\end{eqnarray}
where the phase space factor $I_2(E_\pi^{\rm max})$ is defined as
\begin{widetext}
\begin{eqnarray}\label{I2no}
I_2(E_\pi^{\rm max}) = \frac{\int dM_{K\pi}^2 \mbox{BW}^*_{K^*}(M_{K\pi})
\frac{E_\gamma}{E^{(0)}_\gamma} 
\int^{E_\pi^{\rm max}}\! dE_\pi \frac{|\vec p^\perp_\pi\,|^2}{E_\pi+\Delta}}
{\int dM_{K\pi}^2 |\mbox{BW}_{K^*}(M_{K\pi})|^2 
\int^{E_\pi^{\rm max}}\! dE_\pi |\vec p^\perp_\pi\,|^2}  &=& 0.20 + 0.11 i
\mbox{ GeV}\,, 
\end{eqnarray}
\end{widetext}
and the numerical value corresponds to $E_\pi^{\rm max} = 0.5 \text{~GeV}$.
As mentioned, for pion energies above $1.0-1.5$ GeV, 
the right-handed photon amplitude is power suppressed, so it can be expected 
to be numerically small. We estimate the contribution from the 
hard-soft region 0.5 GeV $< E_\pi < 1.0-1.5$ GeV by assuming the validity of the low
energy expression for the decay amplitudes over this range.
Taking $E_\pi^{\rm max}  = 1.5$ GeV 
replaces the numerical value in Eq.~(\ref{I2no}) with $I_2 = 0.14 + 0.05 i$ GeV.
In both cases discussed above, the contribution of the nonresonant
(third) term in the braces in Eq.~(\ref{Stot}) is less than $0.5\%$,
and is thus negligible.

We neglected in this computation the presence of higher $K^*$ resonances.
The Particle Data Book \cite{PDG} lists four $K^*$ resonances 
in the region $m_K + m_\pi \leq M_{K\pi} \leq 1.8$ GeV, which can appear in
the $K_S\pi^0$ invariant mass spectrum  (with quantum numbers $J^P=1^-,2^+,3^-,\dots$). 
Their inclusion does not change the leading order nonresonant right-handed 
photon amplitude computed here, but introduces additional power suppressed 
effects similar to those parameterized by $(h_s,\phi_s)$. In the narrow width
limit, their effect is to replace $h_s\cos\phi_s$ in Eq.~(\ref{Stot}) with
\begin{eqnarray}
h_s\cos\phi_s \to \frac{1}{1+\sum_i x_i}
(h_s\cos\phi_s + \sum_i x_i h_s^i\cos\phi_s^i)
\end{eqnarray}
with $x_i = Br(B\to K^*_i\gamma) Br(K^*_i\to K\pi)/Br(B\to K^*\gamma)$,
and $(h_s^i, \phi_s^i)$ new parameters for $B\to K^*_i\gamma$ defined 
analogously to Eq.~(\ref{HRLmodel}). Of the four kaon resonances contributing to the
sum, only two of them decay into $K\pi$ with a branching fraction 
larger than 30\%: $K_2^*(1430)$ and $K^*(1680)$, with $x_i = 0.15$ and 0.01
respectively. This shows that the contributions of the higher kaon resonances
are likely very small and can be neglected.

Our results demonstrate that the nonresonant contribution to the mixing-induced
CP asymmetry is negligibly small, and the SM contamination is dominated by
the right-handed photon amplitude in $\bar B\to K^*\gamma_R$, parameterized
by $h_s \cos\phi_s$. Our results show that 
averaging the nonresonant and resonant
measurements of the $S$ parameter is a justified procedure. 

\vspace{0.5cm}

\section{V. Conclusion}

We studied in this paper the Standard Model prediction for the
mixing-induced CP asymmetry parameter in $B^0\to K_S\pi^0\gamma$ decay.
This decay is important as a probe for new physics manifested through
a right-handed photon in $b\to s\gamma$ decay. The naive expectation
\cite{Atwood:1997zr}
for the $S$ parameter in the SM is $S = -2\sin 2\beta (m_s/m_b) \sim 2\%$.
We computed the corrections to this prediction introduced
by strong interaction effects.

In the kinematical region with $M_{K_S\pi^0} \sim M_{K^*}$ 
and a soft pion in the rest frame of the B meson, 
there is a unique SM mechanism contributing to the $S$ parameter at leading
order in $\Lambda/m_b$,
arising from the $B^*$ pole diagrams. These effects are factorizable
and calculable using a combination of SCET and heavy hadron
chiral perturbation theory \cite{GPchiral}. In addition, power suppressed
effects can introduce a potentially sizeable contamination from nonfactorizable
graphs with the photon coupling to the charm quark loop~\cite{pol}. These
are difficult to compute in a reliable way, and a simple power counting
estimate allows a right-handed photon amplitude as large as $\sim 9\%$.

We performed a detailed numerical study of the nonresonant effects.
We find that the leading order $B^*$ pole effect is numerically small.
It introduces a weak dependence of the CP asymmetry $S_{K_S\pi^0\gamma}$ on 
$M_{K\pi}$. The dominant SM contamination is from power suppressed
effects in the $\bar B\to K^*\gamma_R$ resonant amplitude, and our best 
estimate in the $M_{K\pi}$ dependent asymmetry is $|S_{K_S\pi^0\gamma}^{\rm SM}| 
\leq 8\%$ (see Fig.~4). 

When integrated over $M_{K\pi}$, the nonresonant effect is 
practically negligible, and the CP asymmetry $S_{K_S\pi^0\gamma}$ is
dominated by the resonant $\bar B\to \bar K^*\gamma_R$ amplitude.
This means that averaging the results of the resonant and nonresonant
measurements, as currently done at B factories, is a justified procedure.
If improved measurements of the CP asymmetry confirm the present average
$|S| \sim 8\%$, this would be consistent with a power suppressed
correction in the SM. We reiterate that the naive estimate
$S \sim -2 (m_s/m_b) \sin 2\beta$ seriously underestimates the
value of the $S$ parameter in the SM.
Furthermore, one would also expect the agreement
between resonant and nonresonant measurements to improve. 
\vspace{0.5cm}


%


\begin{acknowledgments}
We thank Yuval Grossman and Zoltan Ligeti for collaboration
on related work and comments. DP thanks Jure Zupan for discussions and
advice.
This work was supported in part by the DOE under grant DE-FG03-97ER40546 (BG),
and by the DOE under cooperative research agreement
DOE-FC02-94ER40818 and by the NSF under grant PHY-9970781 (DP).
\end{acknowledgments}

\end{document}